\documentclass{aa}
\pdfoutput=1
\usepackage{wasysym}
\usepackage[varg]{txfonts}
\usepackage{natbib}
\usepackage{graphicx,subcaption,xfrac,multirow,color}
\usepackage{hyperref}
\bibpunct{(}{)}{;}{a}{}{,}

\newcommand{\mps}{\mathrm{m}/\mathrm{s}}
\newcommand{\f}[2]{\mathrm{#1}\left(#2\right)}
\newcommand{\e}[1]{\cdot10^{#1}}

\begin{document}

\title{Formation of the Janus-Epimetheus system through collisions}
\abstract {Co-orbital systems are bodies that share the same mean orbit. They can be divided into different families according to the relative mass of the co-orbital partners and the particularities of their movement. Janus and Epimetheus are unique in that they are the only known co-orbital pair of comparable masses and thus the only known system in mutual horseshoe orbit.}{We aim to establish whether the Janus-Epimetheus system might have
formed by disruption of an object in the current orbit of Epimetheus.}{We assumed that four large main fragments were formed and neglected smaller fragments. We used numerical integration of the full N-body problem to study the evolution of different fragment arrangements. Collisions were assumed to result in perfectly inelastic merging of bodies. We statistically analysed the outcome of these simulations to infer  whether co-orbital systems might have formed from the chosen initial conditions.}{Depending on the range of initial conditions, up to $9\%$ of the simulations evolve into co-orbital systems.
Initial velocities around the escape velocity of Janus yield the highest formation probability.
Analysis of the evolution shows that all co-orbital systems are produced via secondary collisions.
The velocity of these collisions needs to be low enough that the fragments can merge and not be destroyed.
Generally, collisions are found to be faster than an approximate cut-off velocity threshold. However, given a sufficiently low initial velocity, up to 15\% of collisions is expected to result in merging.
Hence, the results of this study show that the considered formation scenario is viable.} {}
\author{Lucas L. Treffenstädt\inst{\ref{unesp},\ref{ubt}}
    \and D\'ecio C. Mour\~ao\inst{\ref{unesp}}
        \and Othon C. Winter\inst{\ref{unesp}}}
\institute{Univ Estadual Paulista – UNESP, Grupo de Din\^amica Orbital \& Planetologia, Guaratinguet\'a, CEP 12516-410, SP, Brazil\label{unesp} \and Theoretische Physik II, Physikalisches Institut, Universität Bayreuth, D-95440 Bayreuth, Germany\label{ubt}}
\titlerunning{Formation of the Janus/Epimetheus System through Collisions}
\date{received date / accepted date}

\maketitle

\section{Introduction}
In a co-orbital system, two or more bodies share a mean orbit around a central body in a 1:1 orbital resonance.
The possible orbits have been described by Lagrange in his \emph{Analytical Mechanics} in 1788.
We speak of tadpole orbits when one body librates around the Lagrange points L$_4$ or L$_5$ of another.
Libration around L$_3$, L$_4,$ and L$_5$ constitutes a horseshoe orbit \citep{brown1911}.
Many trojan asteroids are in co-orbital trajectories with planets, but only in the Saturnian system are satellites known to be in co-orbital motion. 
While there are many examples of bodies with much smaller co-orbital partners, the Saturn moons Janus and Epimetheus are a unique system of two bodies with similar mass in horseshoe orbit.
This co-orbital motion  was confirmed in 1980 by the Voyager I probe \citep{synnott1981orbits}.  
\citet{harrington1981dynamics} showed the stability of this system through numerical integrations including the oblateness of Saturn.
\citet{dermott1981dynamics} developed an analytical study estimating the radial and angular amplitudes of each horseshoe orbit.
A theory of motion was proposed two years later by \citet{yoder1983theory}, who explored the phase space of co-orbital systems with comparable masses, identifying the equilibrium points and showing the possible trajectory types (tadpole and horseshoe). 
Considering a system with up to nine equal-mass co-orbital bodies, \citet{salo1988} found the location of the equilibrium points and their stability. Following a Hamiltonian approach, \citet{sicardy2003} demonstrated the connection between co-orbital systems of comparable masses with that of the restricted case.

The formation of co-orbital systems has been the subject of intensive studies.
In 1979, \citet{yoder1979origins} proposed three possible mechanisms for the formation of the Jovian trojans:
\begin{enumerate}
\item early capture during the accretion of Jupiter,
\item the continuous capture of asteroids, and
\item transfer of Jovian satellites to the Lagrange points.
\end{enumerate}
In the first scenario, gas drag plays an important role.
\citet{peale1993trojans} and \citet{murray1994stability} showed that, in the presence of a gas, L$_4$ and L$_5$ can still be stable for proto-Jupiter.
In 1995, \citet{kary1995gasdrag} showed that a protoplanet on an eccentric orbit can indeed trap planetesimals in a 1:1 resonance, decaying from a large-libration horseshoe orbit to a tadpole orbit around L$_5$.
\citet{fleming2000} explored the effect due to mass accretion and orbital migration on the dynamics of co-orbital systems. They showed how these perturbations affect the radial and angular amplitudes of the trajectories.
\citet{kortenkamp2005capture} later proposed a mechanism to efficiently capture planetesimals on a satellite orbit after gravitational scattering by the protoplanet, making this a more likely fate than trapping in a co-orbital.
\citet{chanut2008nebular} determined the size limit of trapping planetesimals via nebular gas drag to be 0.2m per AU of orbit of the protoplanet.
They \citep{chanut2013cavity} later extended their approach to a non-uniform gas with a cavity in the vicinity of a heavy planet, but found little to no co-orbitals for eccentricities comparable to Janus and Epimetheus.
All these scenarios result in co-orbital configurations with a large secondary body and a small partner, that is, with a low mass ratio. This almost excludes trapping as a mechanism for the formation of the Janus-Epimetheus system.
Furthermore, \citet{hadjidemetriou2011gasdrag} showed that a system of two equal-mass bodies in a 1:1 resonance tends to evolve under gas drag into a closely bound binary satellite system.
Another take on the subject by \citet{cresswell2006gasdrag} was the study of the evolution of a swarm of similarly sized planetesimals in a gas disk.
They showed that the formation of mean-motion resonances, including 1:1 resonances, is commonplace, but that the bodies are still prone to be lost to accretion by the central body.
\citet{giuppone2012cavity} followed a similar approach to \citet{chanut2013cavity}, but studying a hypothetical extrasolar system.
They showed that a pair of inwardly migrating planets can enter into a 1:1 resonance at the edge of a cavity in a gas disk. Whether their result is applicable to Janus and Epimetheus is not known to us.

\citet{laughlin2002collision} showed that stable 1:1 resonances of equal-sized planets around a central body exist.
They hypothesized that at the L$_5$ point of a planet in the accretion disk of a protostar, another planet could accrete.
\citet{beauge2007accretion} investigated formation by accretion of a swarm of planetesimals in a tadpole orbit around L$_4$ or L$_5$ of a giant extrasolar planet.
They showed that an Earth-type body can form, but with a very low efficiency and thus a low final mass. 
\citet{chiang2005neptune} considered this a likely process for the formation of the Neptune trojans. 
\citet{izidoro2010congential} studied the mechanism for the Saturnian system and showed that formation of co-orbital satellite pairs with a relative mass of $10^{-13}$ to $10^{-9}$ is possible.
This again rules out the formation of a co-orbital pair with similar masses.

While much research has been done on the subject of co-orbital systems, there is as yet no satisfactory explanation for the origin of the Janus-Epimetheus system in particular.
Our approach here is based on the collision of two bodies at or close to the current orbit of Epimetheus.
This is consistent with the equal densities (see Table~\ref{tab:jeproperties}) and compositions of Janus and Epimetheus.
Moreover, results by \citet{charnoz2010formation} show a recent formation of the Saturn moonlets from its rings through series of subsequent collisions. \citet{crida2012formation} later extended this scenario to all regular moons of Saturn. In this context, low-velocity collisions are frequent and are almost inevitable, especially for the
predecessors of Janus and Epimemetheus.
We assume the disruption into four main fragments that later collide again constructively to form a pair of bodies in a co-orbital configuration.
We establish the statistical likelihood of this outcome in 8000 numerical simulations.
We used Chambers' \emph{Mercury} \citep{chambers1999mercury} N-Body integrator with a Burlisch-Stoer algorithm.

In Sect. 2, we explain our model in detail and introduce the methods used to obtain our results. Section 3 shows the setup and results of our simulations. Finally, we discuss the viability of our model considering cases with six and eight fragments, and propose opportunities for future studies in Sect. 4.

\section{Model and method}

\newlength{\panelwidth}
\setlength{\panelwidth}{0.28\textwidth}
\newlength{\subfigwidth}
\setlength{\subfigwidth}{0.38\textwidth}
\begin{figure*}
\centering
\begin{subfigure}{\subfigwidth}
\centering
\includegraphics[width=\panelwidth]{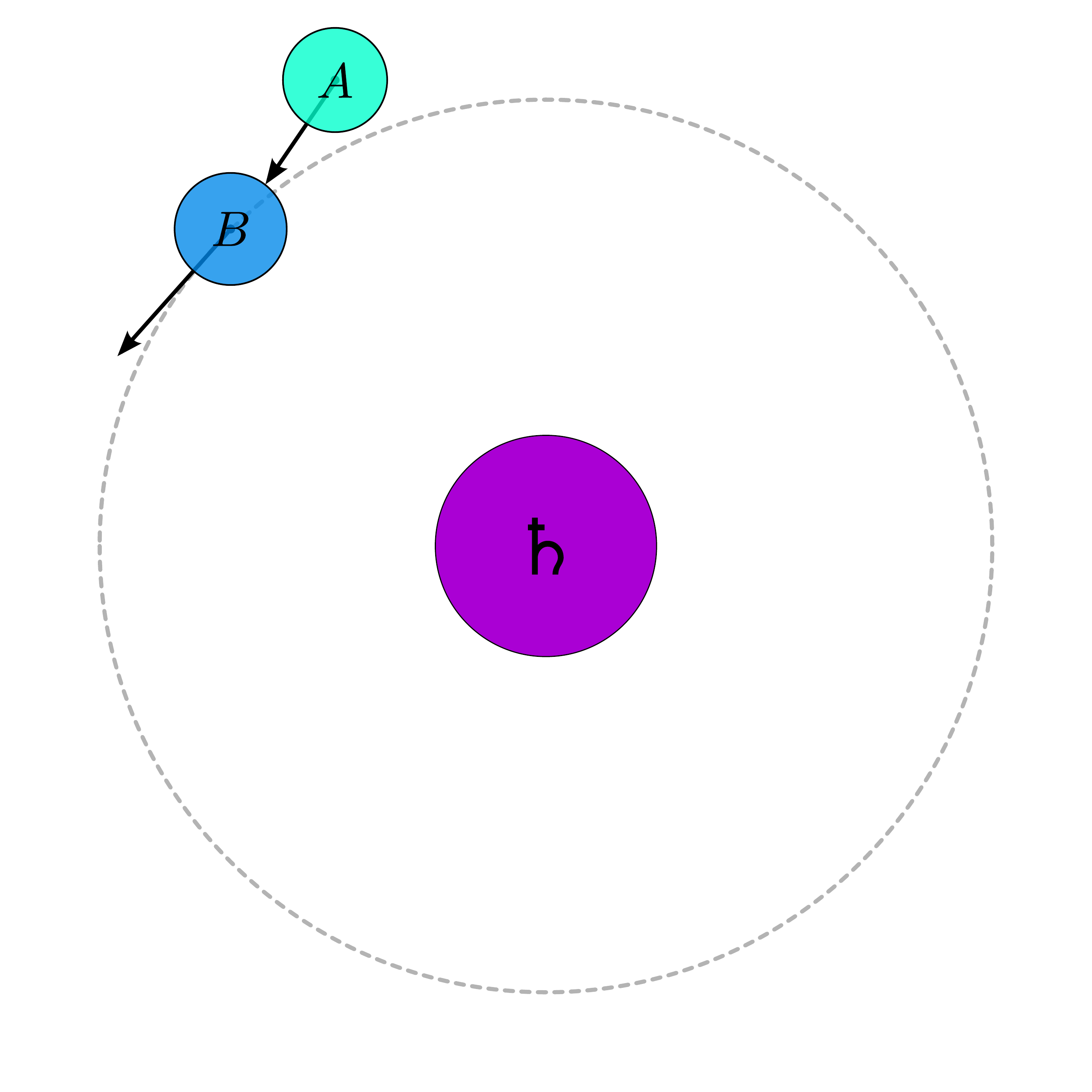}
\caption{Collision event. Body $A$ collides with body $B$ on Epimetheus' orbit.}
\label{fig:scenario1}
\end{subfigure}~
\begin{subfigure}{\subfigwidth}
\centering
\includegraphics[width=\panelwidth]{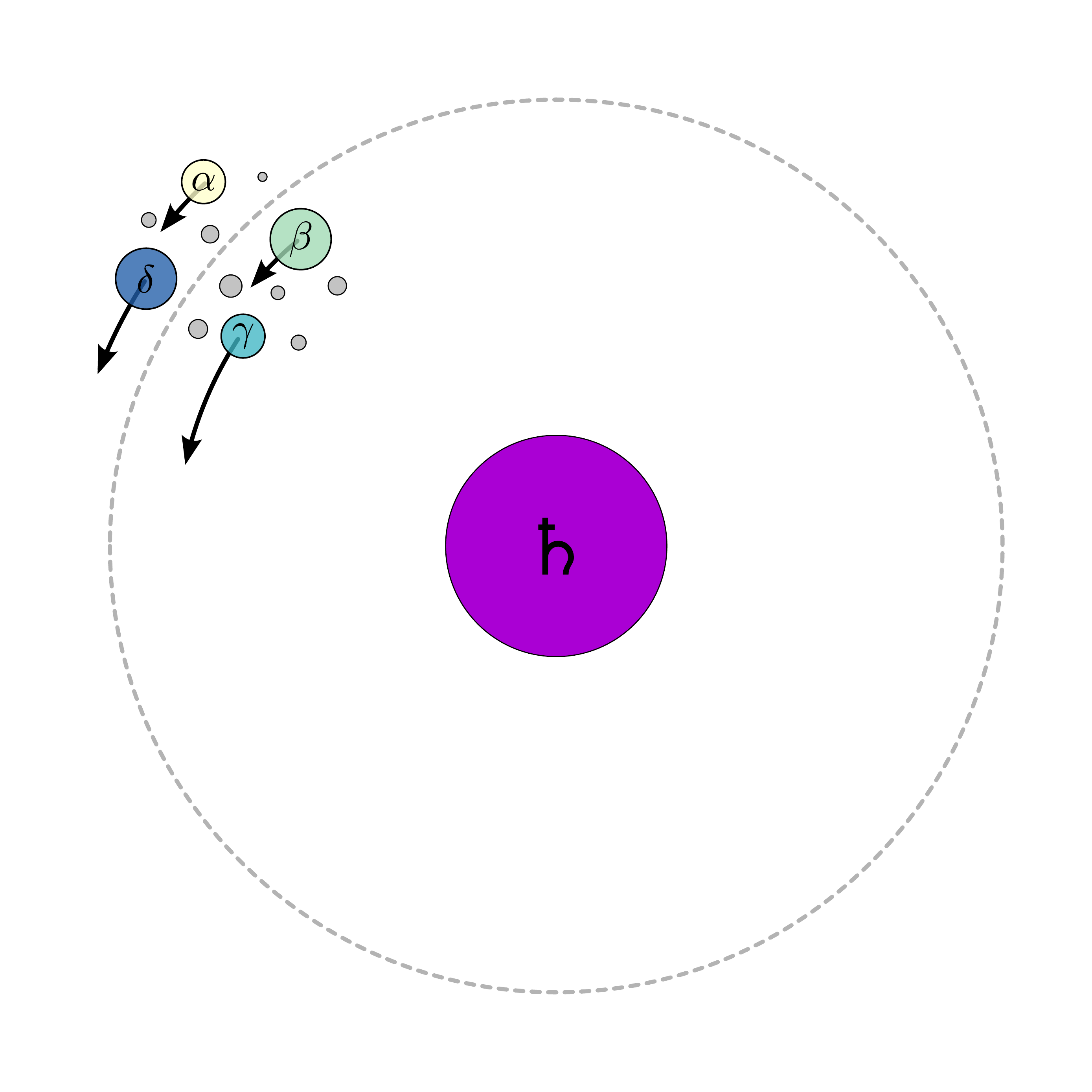}
\caption{Disruption into four main fragments. The remaining debris is assumed to be negligible.}
\label{fig:scenario2}
\end{subfigure}

\begin{subfigure}{\subfigwidth}
\centering
\includegraphics[width=\panelwidth]{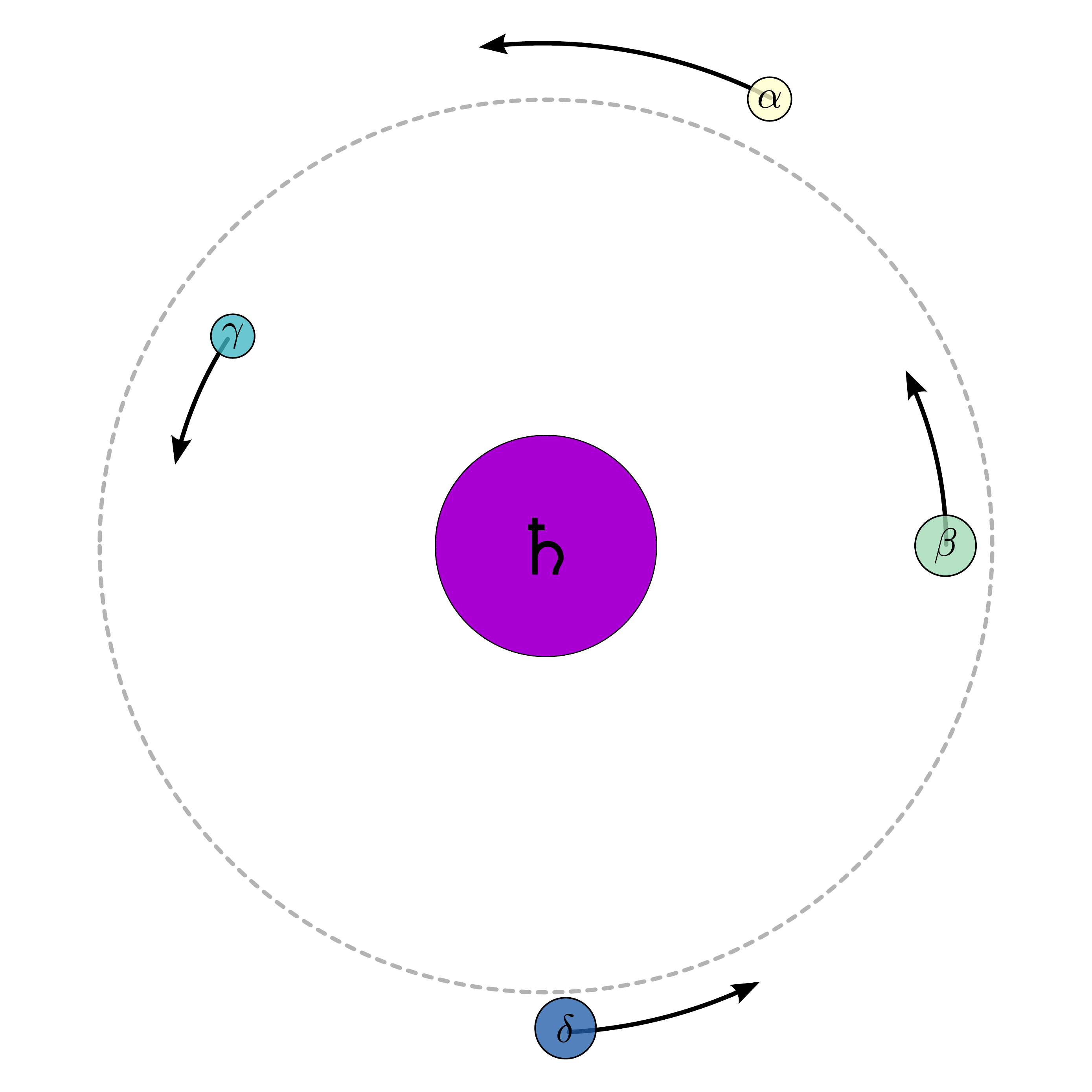}
\caption{The main fragments distributed on orbit.}
\label{fig:scenario3}
\end{subfigure}~
\begin{subfigure}{\subfigwidth}
\centering
\includegraphics[width=\panelwidth]{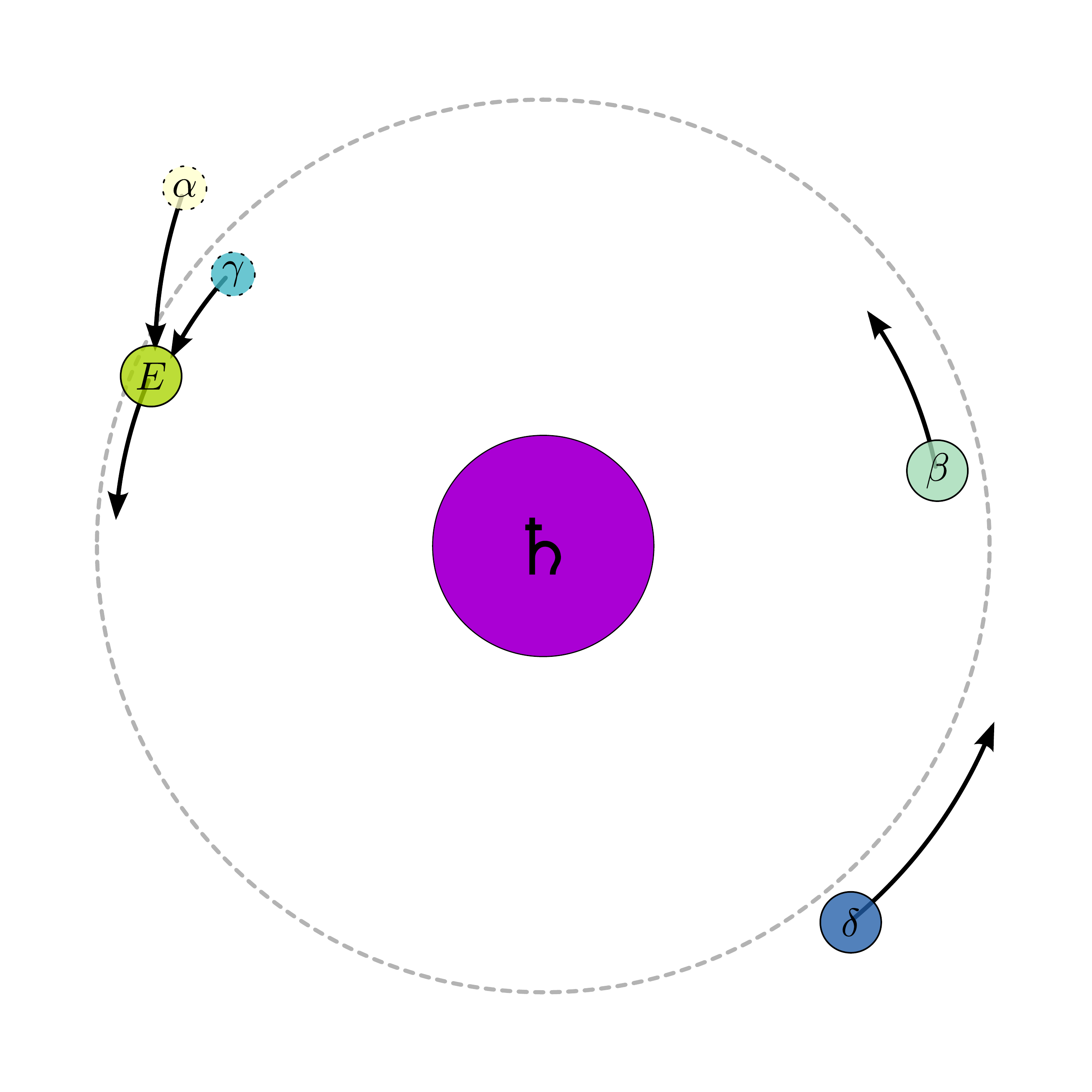}
\caption{Fragments collide to form Epimetheus.}
\label{fig:scenario4}
\end{subfigure}

\begin{subfigure}{\subfigwidth}
\centering
\includegraphics[width=\panelwidth]{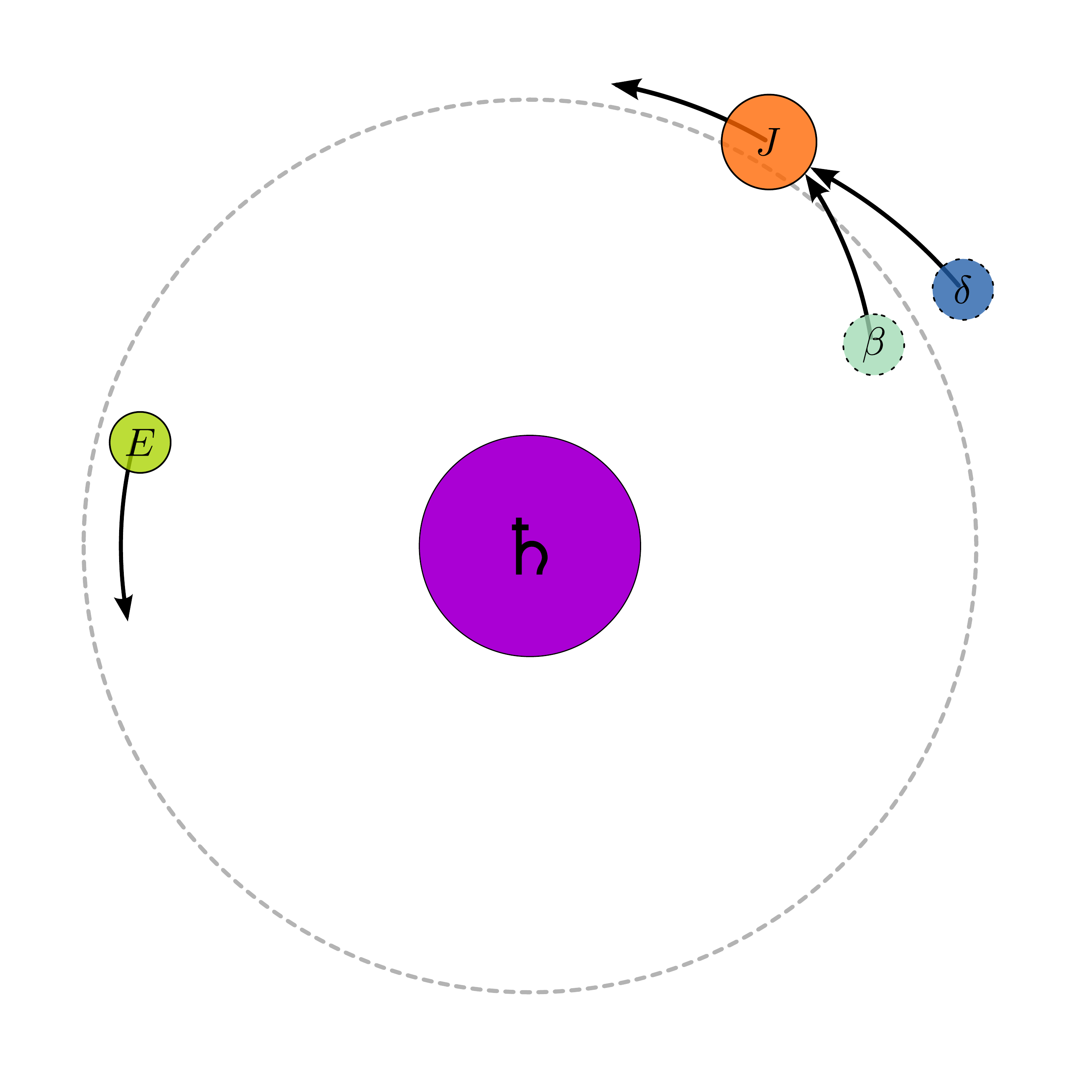}
\caption{Fragments collide to form Janus}
\label{fig:scenario5}
\end{subfigure}~
\begin{subfigure}{\subfigwidth}
\centering
\includegraphics[width=\panelwidth]{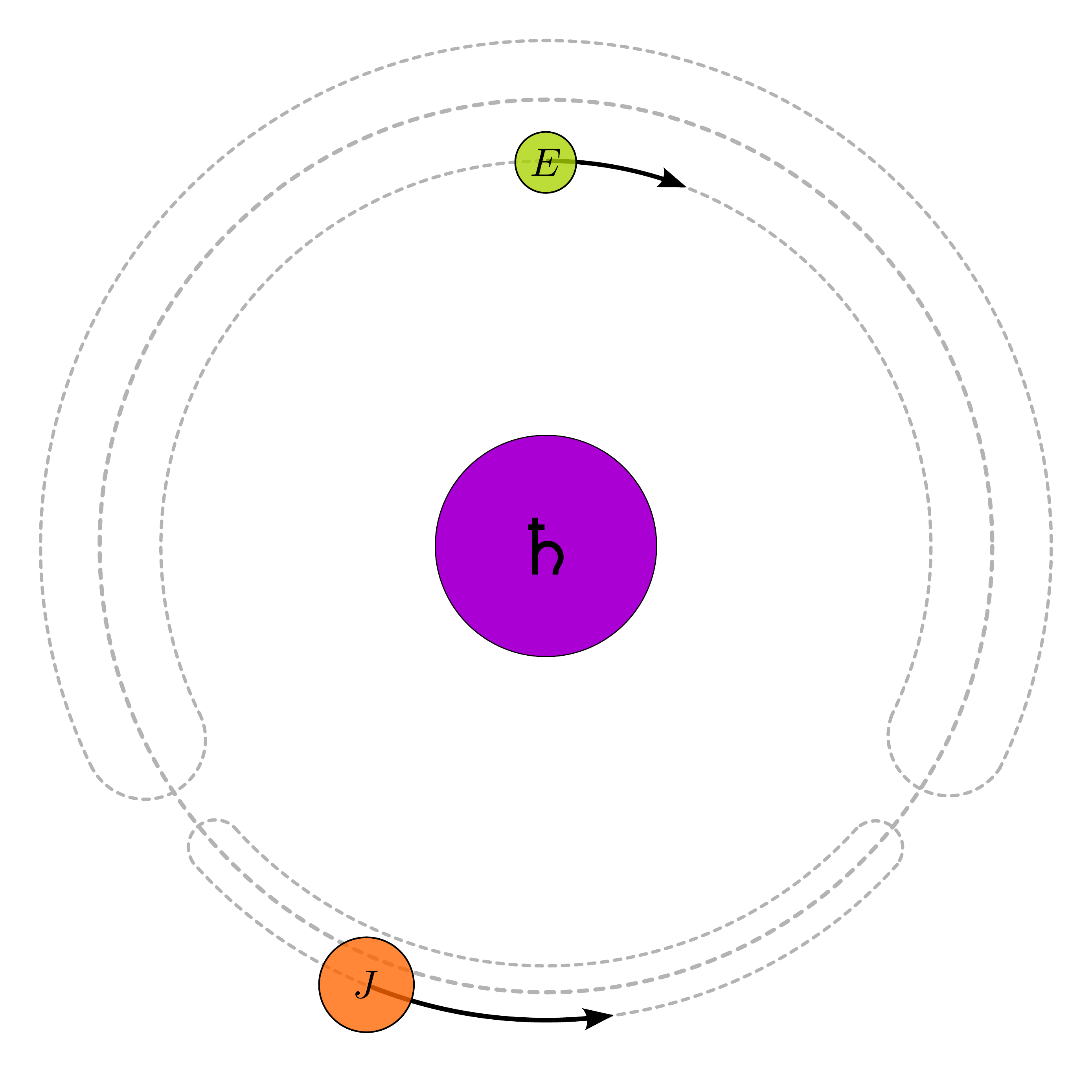}
\caption{Janus and Epimetheus enter into horseshoe orbits.}
\label{fig:scenario6}
\end{subfigure}
\caption{Sketch of typical co-orbital formation scenario from our simulations.}
\label{fig:scenario}
\end{figure*}

\begin{figure}
        \centering
        \includegraphics[width=0.6\linewidth]{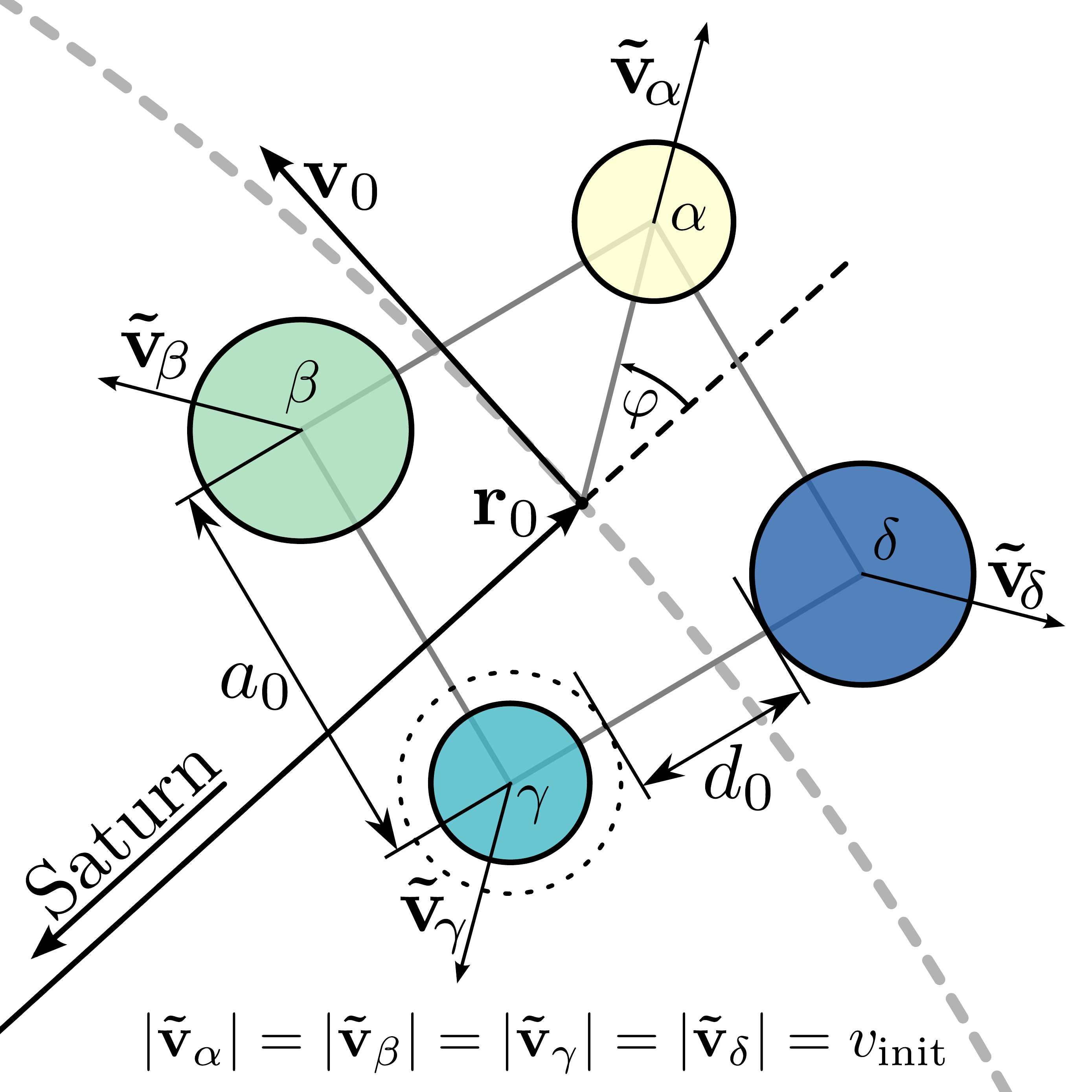}
        \caption{Initial configuration of the simulations.
        $\mathbf{\tilde{v}}_i$ are given in the reference frame moving with $\mathbf{v}_0$.
    The initial velocity of each fragment in the simulation frame of reference is given by $\mathbf{v}_i$ = $\mathbf{\tilde{v}}_i + \mathbf{v}_0$.
    The direction of $\mathbf{\tilde{v}}_i$ is given by the vector from the centre of the square to the respective vertex.
    The collision fragments $\alpha$, $\beta$, $\gamma,$ and $\delta$ are positioned on the vertices of a square centred on $\mathbf{r}_0$, the position vector with respect to Saturn.
    The angle $\varphi$ gives the rotation of the square around the vector $\mathbf{r}_0 \times \mathbf{v}_0$ (the normal of the orbital plane).
    For $\varphi = 0$ $\alpha$, $\gamma,$ and Saturn are connected by a straight line, with $\gamma$ between $\alpha$ and Saturn.
    The fragments are always set on the vertices in a counter-clockwise sense, as shown in the sketch.
    The dimensions here are not to scale.}
        \label{fig:initial}
\end{figure}

We investigated whether the Janus/Epimetheus co-orbital system might have formed by disruption of a large body in the mean orbit of Epimetheus.
Figure~\ref{fig:scenario} [a,b] illustrates the scenario that we believe can lead to co-orbital configurations.
We assumed that four large fragments are formed by a collision event.
While the initial collision is not subject of this study, we note that a low collision velocity and low density (compare Table~\ref{tab:jeproperties}) of the colliding bodies is a requirement.
The collision is neither completely destructive nor is it constructive.
After the collision event, fragments spread out on orbits around Saturn.
Fragments can be lost by ejection from the system, collision with Saturn, or collision with each other.
In the last case, fragments can unite to form a larger body, given a sufficiently low collision velocity \citep[see][]{kortenkamp2000collision}.

Our simulations start right after the initial collision event, with the four fragments $\alpha$, $\beta$, $\gamma,$ and $\delta$ on the vertices of a square, moving in opposite directions with equal speed (see Fig.~\ref{fig:initial} for details).
We always chose equal masses for bodies opposite on the square.
The length $a_0$ of the sides of the square was chosen to be $a_0 = 2R_1+d_0$ , where $R_1$ is the radius of the largest body and $d_0=100\mathrm{m}$.
The parameters varied in this study include the masses $m_i$ of the fragments (given as fractions of the mass of Janus and
Epimetheus, see Tables~\ref{tab:jeproperties} and~\ref{tab:parameters}), the orientation of the square in the orbital plane denoted by the angle $\varphi$ (going from $0$ to $\pi$ because of symmetry), and the initial speed $v_\mathrm{init}$ (see Fig.~\ref{fig:initial}) of the fragments in the coordinate system moving with $\mathbf{v}_0$ relative to Saturn.
The position and velocity $\{\mathbf{r}_0,\mathbf{v}_0\}$ of the centre of the square are equal to the state vector of Epimetheus at an arbitrarily chosen point of its orbit. This is the same for all simulations.

\begin{table*}
        \centering
    \caption{Parameters of Janus and Epimetheus \citep{jacobson2008orbits,thomas2013details}}
    \label{tab:jeproperties}
    \begin{tabular}{r | l | c r | c r}
        parameter & unit & \multicolumn{2}{c |}{Epimetheus} & \multicolumn{2}{c}{Janus} \\
    \hline
    mass & $m_{\saturn}$ & $m_\mathrm{epi}$ & $0.11667\cdot10^{-8}$ & $m_\mathrm{jan}$ & $0.58333\cdot10^{-8}$ \\
    escape velocity &$\mps$ & & $39$ & $v_\mathrm{esc}$ & $65$ \\
    density & $\mathrm{kg}/\mathrm{m}^3$ & $\rho_\mathrm{epi}$ & $640\pm62$ & $\rho_\mathrm{jan}$ & $630\pm30$ \\
    mean radius & $\mathrm{R}_0$ & & $3.85\cdot10^{-4}$ & & $6.88\cdot10^{-4}$ \\
    semi-major axis & $\mathrm{R}_0$ & $a_\mathrm{epi}$ & 1.000172 & $a_\mathrm{jan}$ & $0.999842$ \\
    eccentricity & & $e_\mathrm{epi}$ & 0.0097 & $e_\mathrm{jan}$ & $0.0068$ \\
    \hline
        \end{tabular}
    
        \tablefoot{$m_{\saturn}$ is the mass of Saturn and $\mathrm{R}_0$ is the mean distance of Janus and Saturn.}
\end{table*}

We used the package \emph{Mercury} to numerically integrate the system using a Burlisch-Stoer algorithm \citep{chambers1999mercury}.
We included the moons Mimas and Tethys into our simulations, since they exert the strongest influence on the Janus-Epimetheus system, but neglected other objects.
We also considered the oblateness of Saturn through J$_2$, J$_4,$ and J$_6$ (Table~\ref{tab:saturn}).
To obtain the state vectors of Mimas, Tethys, and Epimetheus (centre of the square from which the fragments start), we used the data given by \citet{jacobson2008orbits} and
followed the approach presented by \citet{renner2006elements} to transform the geometric elements. 
We took the mean distance between the system Janus-Epimetheus and Saturn as the unit of length $\mathrm{R}_0$.
The simulation code was set to record the state vectors of all bodies every $\sfrac{1}{100}$ day.
We need this high time resolution to study the velocities and angles of collisional events between fragments.
\emph{Mercury} assumes a perfectly inelastic collision where the colliding bodies aggregate into a new body of the added masses of the colliding bodies, and the momentum is preserved.
We therefore needed to analyse the collisions to ensure that
this assumption is realistic.
The outcome of a collision is strongly dependent on the collision velocity and the physical characteristics (density, strength,
etc.) of each body. 
\citet{kortenkamp2000collision} established an escape velocity term
\begin{equation}
\label{eqn:vmax}
v_\mathrm{max}^2 = \frac{2\mathrm{G}\left(m_1+m_2\right)}{R_1+R_2}
\end{equation}
for the encounter of rocky bodies, where $m_i$ is the mass of body $i$, $R_i$ is its radius, and $\mathrm{G}$ is the gravitational constant.
For collision velocities below $v_\mathrm{max}$, a collision can be assumed to be constructive.
For a collision of Janus and Epimetheus, $v_\mathrm{max}\approx 58\mps$.
However, their study involved bodies with a density about five times as high as that of Janus and Epimetheus.
Since in a collision event between fluffy bodies (such as Janus and Epimetheus, whose density is lower than $700\mathrm{kg}/\mathrm{m}^3$) kinetic energy is lost to deformation, higher velocity collision events can still be constructive.
High-velocity collisions can also result in the catastrophic disruption of one or both colliding bodies.
Multiple studies on the disruption threshold \citep{colwell1994disruption,benz1999disruption,leinhardt2009disruption} have been conducted.
Using Fig. 11 from \cite{leinhardt2009disruption} and a target radius of $6\e{4}$m (the rough size of Epimetheus), we obtain a disruption threshold of $Q_\mathrm{D}^{*}\approx 6.5\e{3}$J/kg.
This gives a cutoff velocity
\begin{equation}
v_\mathrm{D}=\sqrt{2Q_\mathrm{D}^{*}\frac{m_\mathrm{T}}{m_\mathrm{P}}}
\end{equation}
(where $m_\mathrm{T}$ is the mass of the target and $m_\mathrm{P}$ the mass of the projectile) of approximately $60\mps$ for a collision with Janus.
Therefore we used the first estimate of $58\mps$ as a cutoff, keeping in mind that the actual threshold might still vary and is in fact dependent on which of the fragments collide.

Because of the large number of simulations, we needed an automatic filtering algorithm to find the systems that evolved into a co-orbital.
We used a four-step filtering process.
\begin{enumerate}
\item We selected the systems in which at least two fragments have survived until the end of the simulation (20-30 years). Fragments can be lost through ejection, collision with Saturn, or collision with each other.
\item From the remaining systems, we selected the systems with exactly one pair of fragments that survived until the end of the simulation and overlap in their range of semi-major axes for the time after the last collision in the simulation.
\item From these, we selected the systems in which the mean semi-major axis of the surviving fragments is smaller than $1.2\mathrm{R}_0$ and the minima and maxima of the semimajor axes are no more than $0.02\mathrm{R}_0$ apart. 
\item We plot the semi-major axes of the remaining fragments against time and inspected them visually to check for periodic crossings.
We thus determined false positives of the automatic filtering process and removed them.
\end{enumerate}

\newlength{\firstcol}
\settowidth{\firstcol}{Janus Mean Distance}
\begin{table}
        \centering
        \caption{Properties of Saturn as used in the simulations}
        \label{tab:saturn}
        \begin{tabular}{r c r l}
        \multirow{3}{\firstcol}{\raggedleft zonal gravitational coefficients}& J$_2$ & $1.62980\cdot10^{-2}$ & \\
        & J$_4$ & $-9.150\cdot10^{-4}$ & \\
        & J$_6$ & $1.030\cdot10^{-4}$ & \\
        mass & $m_{\saturn}$ & $5.688\cdot10^{26}$ & kg \\
        equatorial radius & $r_{\saturn}$ & $0.4$ & $\mathrm{R}_0$ \\
        Janus mean distance & $\mathrm{R}_0$ & $1.51464\cdot10^{8}$ & m \\
        \end{tabular}
\end{table}

\section{Simulations and results}

\begin{figure*}
	\centering
	\input{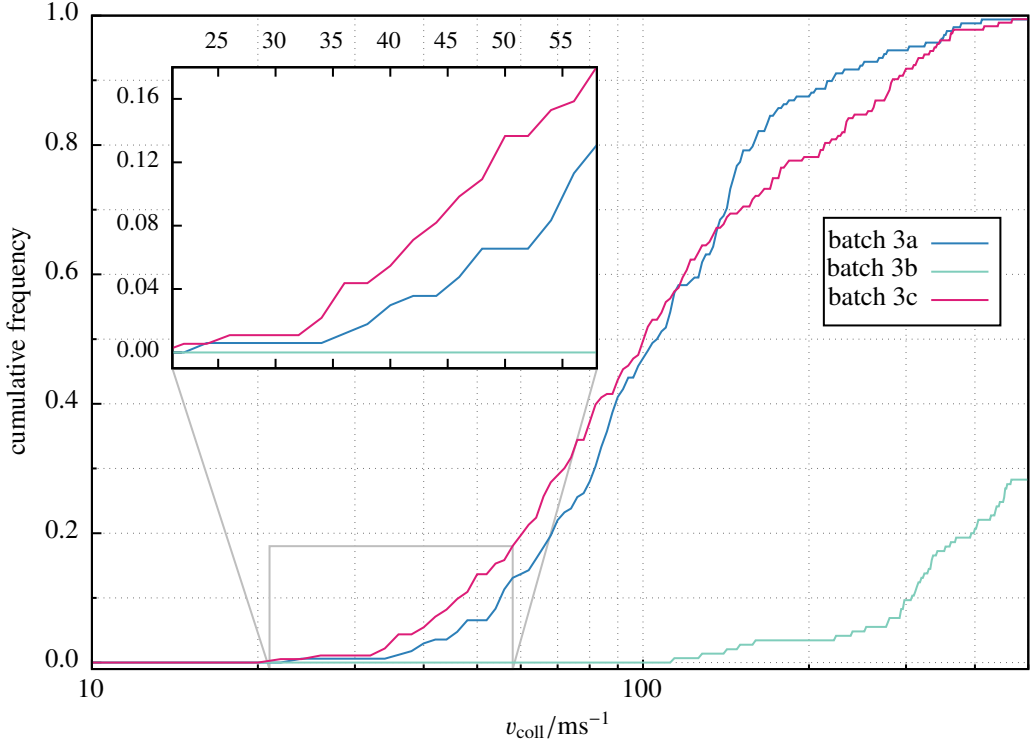}
	\caption{Cumulative distribution of collision velocities for co-orbital systems in batch 3}
	\label{fig:lowmass-comparison}
\end{figure*}

\begin{figure}
	\centering
	\input{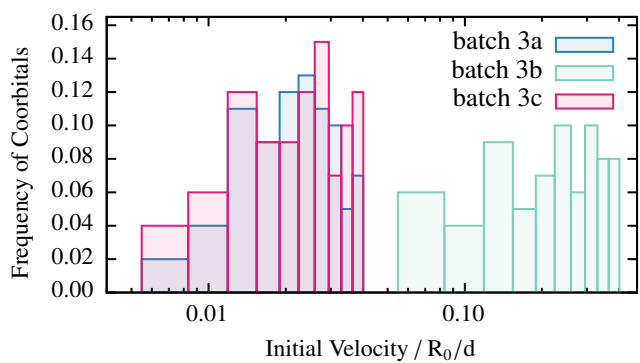}
	\caption{Frequency of co-orbitals as a function of the initial speed $v_\mathrm{init}$ of the fragments for batch 3}
	\label{fig:coorbital-frequency}
\end{figure}

In each simulation, we sampled the parameters $v_\mathrm{init}$ and $\varphi$ in equally spaced steps.
We started our investigation with a batch of 4000 simulations, covering 20 different initial speeds $v_\mathrm{init}$ and 200 different angles $\varphi$ (see Tables~\ref{tab:parameters}-1).
The initial speeds were between 10 and 100 times as high as the escape velocity of Janus.
We found that around $1.5\%$ of the simulations evolved into co-orbital systems.
From the analysis of these systems, we were able to establish a common formation scenario as depicted in Fig.~\ref{fig:scenario}:
After the initial collision event, the fragments first spread out on orbits around Saturn.
Then, they re-collide pairwise, most commonly $\alpha$ with $\gamma$ and $\beta$ with $\delta$.
The velocities of these collision events are very high ($1000\mps$ and above), far beyond where a perfectly inelastic joining, as is assumed by \emph{Mercury}, is realistic.
But the moons created in this way quickly (in just a few orbits) enter into co-orbital horseshoe motion after the second moon has been formed.
No co-orbital systems are the result of ejection of fragments or collision with Saturn.

\begin{table*}
        \centering
        \caption{Parameters for our simulations}
        \label{tab:parameters}
        \begin{tabular}{r | r c l | r c l | c c | c | c | c}
    & \multicolumn{3}{c|}{$v_\mathrm{init} (\mathrm{R}_0/\mathrm{d})$} & \multicolumn{3}{c|}{$\varphi$ (rad)} & \multicolumn{2}{c|}{masses} & \multirow{2}{1.6cm}{\centering integration time (yr)} & \multirow{2}{1.8cm}{\centering frequency of co-orbitals} & \multirow{2}{2.3cm}{\centering collisions below threshold} \\
    set & min & max & steps & min & max & steps & $m_{\alpha / \gamma}$ & $m_{\beta / \delta}$ & & & \\
    \hline
    1 & 0.5 & 4.0 & 20 & 0 & $\pi$ & 200 & $m_\mathrm{epi}$ & $m_\mathrm{jan}$ & 20 & $1.5\%$ & $0\%$ \\
    2 & 0.05 & 0.4 & 10 & 0 & $\pi$ & 100 & $m_\mathrm{epi}$ & $m_\mathrm{jan}$ & 20 & $5.7\%$ & $0\%$ \\
    3a & 0.005 & 0.04 & 10 & 0 & $\pi$ & 100 & $m_\mathrm{epi}$ & $m_\mathrm{jan}$ & 30 & $8.4\%$ & $12.5\%$ \\
    3b & 0.05 & 0.4 & 10 & 0 & $\pi$ & 100 & $m_\mathrm{epi}/2$ & $m_\mathrm{jan}/2$ & 30 & $7.4\%$ & $0\%$ \\
    3c & 0.005 & 0.04 & 10 & 0 & $\pi$ & 100 & $m_\mathrm{epi}/2$ & $m_\mathrm{jan}/2$ & 30 & $8.7\%$ & $15.8\%$ \\
    \hline
        \end{tabular}

        \tablefoot{For constants, see Table~\ref{tab:jeproperties}}
\end{table*}

Then, we investigated whether lower initial velocities (Tables~\ref{tab:parameters}-2) result in lower collision velocities.
We reduced the parameter range of $v_\mathrm{init}$ by a full order of magnitude.
This indeed reduced the collision velocities by nearly an order of magnitude, but not to the point where simple merging of bodies was a realistic outcome for any of the collisions.
We note, however, that co-orbital formation increased to $5.7\%$.

Our third batch of simulations was divided into three parts: First, we simulated with the same masses as before, but at even lower initial velocities (Tables~\ref{tab:parameters}-3a). As expected, the collision velocities decrease even more (see Fig.~\ref{fig:lowmass-comparison} for a comparison of collision velocities).
The portion of co-orbitals increases to $8.4\%$.
Another result is that the frequency of co-orbital formation decreases for very low initial velocities (below $\sim 0.015 \mathrm{R}_0/\mathrm{d}$ or around half of the Janus escape velocity).
The highest observed frequency of co-orbitals is around $0.025 \mathrm{R}_0/\mathrm{d}$.

For the second and third part of this batch, we decreased the masses of the fragments by a half.
One part was ran at initial velocities like in the previous batches (Tables~\ref{tab:parameters}-3b) and the other at lower ones (Tables~\ref{tab:parameters}-3c).
By comparing simulations 3a and 3c, we see that the mass of the fragments has little effect on the distribution of generated co-orbital systems.
Only the decrease in occurrence of co-orbitals for the lowest initial velocities (see Fig.~\ref{fig:coorbital-frequency}) is less prominent for lower mass.
The decrease at these initial velocities indicates a lower limit for the formation of co-orbitals.
Supplementary simulations indicate that for even lower velocities $v_\mathrm{init} < 2\e{-5} \mathrm{R}_0/\mathrm{d}$, formation of co-orbitals is quite unlikely ($\sim0.7\%$) compared to the other simulations.
A cut-off would probably depend on the mass of the fragments, assuming it is related to the escape velocity of the fragments.
This could explain the higher formation probability for lower-mass fragments at low initial velocities.

\begin{figure}
	\centering
	\input{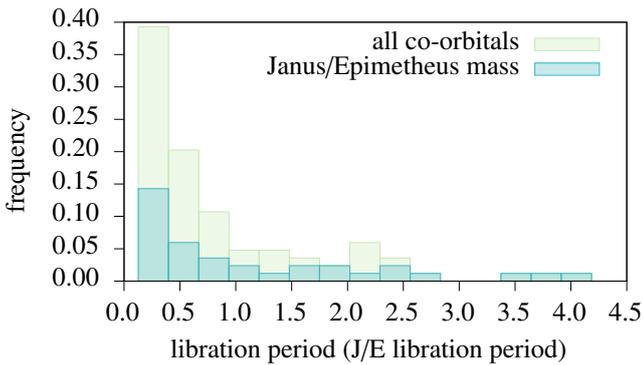}
	\caption{Period of the horseshoe orbits. Note the similar distribution for all masses versus selected masses. The masses of the bodies have almost no influence on the period.}
	\label{fig:period}
\end{figure}

\begin{figure}
	\centering
	\input{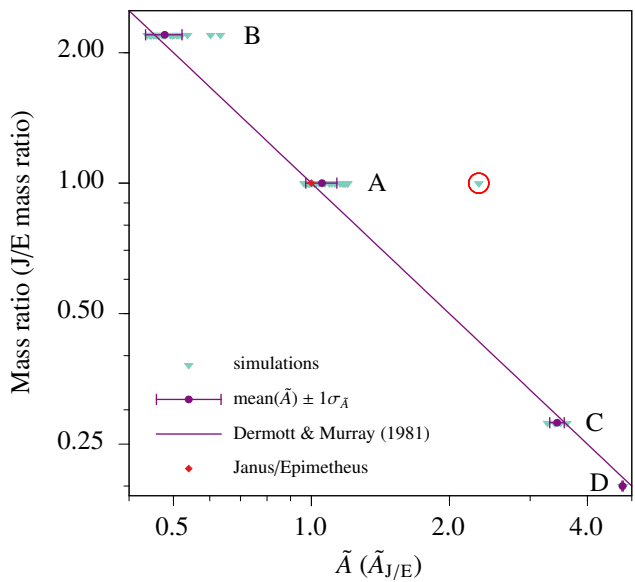}
	\caption{Four classes of co-orbitals by mass ratio and $\tilde{A}$ (see Table~\ref{tab:classes})
	The variables are inversely proportional for horseshoe orbits \citep[][Eqn. (1)]{dermott1981dynamics}.
	The red circle indicates the only pair in a tadpole orbit we observed in this batch of simulations.}
	\label{fig:classes}
\end{figure}

We thus have two constraints on the initial conditions if we wish to determine a range likely to spawn co-orbitals:
The collision velocities are highly dependent on the initial velocities.
Low collision velocities are required for the fragments to collide constructively, thus low initial velocities.
At the same time, the formation frequency of co-orbitals decreases after a certain point, so the initial velocity cannot be arbitrarily low.
The existence and width of a sweet spot for the formation of co-orbitals is key to decide the viability of this model.

We examined the results of simulation batch 3c more closely.
In many but not all cases, the fragments collided in such a way that the final mass of the bodies was equal to Janus and Epimetheus.
This is the case for 36 of the 87 co-orbital systems we obtained in this batch, or $3.6\%$ of all simulations in the batch (see Table~\ref{tab:classes}).
The collision velocities reach up to about 450$\mps$.
The collisions occur at angles smaller than 5\textdegree{} for nearly all cases.
Assuming a limit of $58\mps$ for constructive collision, we find that $15.8\%$ of all collision events in this batch remain (see Fig.~\ref{fig:lowmass-comparison}).
By multiplying the original success rate of $3.6\%$ by $15.8\%$ squared (since both collision events need to be below the threshold), we obtain a corrected success rate of $0.09\%$ to generate Janus and Epimetheus or $0.2\%$ to generate any co-orbital.

We compared the libration period of the horseshoe orbits to the period of the horseshoe orbit of Janus and Epimetheus (see Fig.~\ref{fig:period}).
Janus and Epimetheus cross orbits roughly every four years.
This period is highly variable for different simulations and shows no particular connection to the mass of the moons.
Instead, there seems to be a general preference for short periods.

We see more influence of the mass of the fragments on the trajectories by examining the relative range of the semi-major axes.
With our setup of four fragments, two each of identical mass, there are four possible combinations of a final co-orbital pair, listed in Table~\ref{tab:classes}.

\begin{table*}
        \centering
    \caption{Classes of co-orbital systems by mass. The frequency is relative to all co-orbital systems in simulation batch 3c. Check Fig.~\ref{fig:evolution-examples} for an example orbit for each class defined here.}
    \label{tab:classes}
    \begin{tabular}{c | c c | c c | c | c}
    class & \multicolumn{4}{c |}{combinations} & mass ratio & frequency \\
    \hline
    A & \multicolumn{2}{c}{$(\alpha + \gamma)$} & \multicolumn{2}{c|}{$(\beta + \delta)$} & $4.00$ & $37.5\%$ \\
    \hline
    B & $(\alpha)$ & $(\beta + \gamma + \delta)$ & $(\gamma)$ & $(\alpha + \beta + \delta)$ & $0.11$ & $47.2\%$ \\
    \hline
    C & $(\beta)$ & $(\alpha + \gamma + \delta)$ & $(\delta)$ & $(\alpha + \beta + \gamma)$ & $1.40$ & $6.94\%$ \\
    \hline
    D & $(\alpha + \beta)$ & $(\gamma + \delta)$ & $(\alpha + \delta)$ & $(\beta + \gamma)$ & $1.00$ & $8.3\%$ \\
    \hline
    \end{tabular}
\end{table*}

We now consider the ratio of radial amplitudes
\begin{equation}
        \label{eqn:atilde}
        \tilde{A} \equiv \frac{\f{max}{a_1}-\f{min}{a_1}}{\f{max}{a_2}-\f{min}{a_2}}
,\end{equation}
where $a_1$ is the set of semi-major axes of the heavier body of the final co-orbital pair after the last collision event in the simulation, and $a_2$ is the same set for the lighter body.
For the Janus-Epimetheus system, this is
\begin{equation}
        \tilde{A}_\mathrm{J/E} = 0.2102
.\end{equation}
By plotting $\tilde{A}$ against the ratio of masses (see Fig.~\ref{fig:classes}),
we recover the expected inversely linear correlation between the two.
Each class has similar ratio of radial amplitudes.
There is one simulation with significant deviance (see Fig.~\ref{fig:classes}, red circle) from this law.
Closer investigation reveals that in this case, the surviving bodies enter into a tadpole orbit (see Fig.~\ref{fig:tadpole}). We have found no other case of tadpole orbits in our simulations.

\begin{figure*}
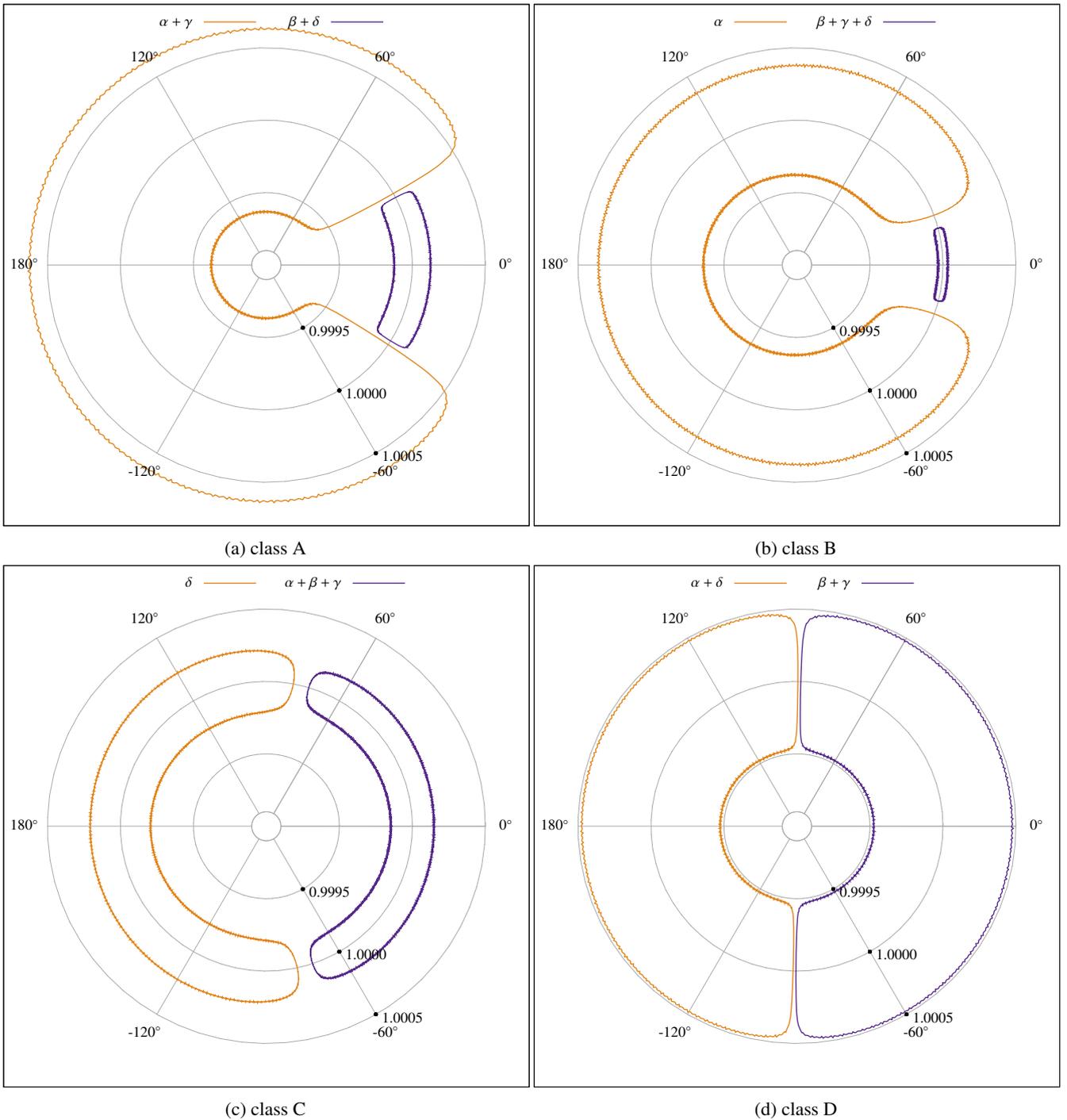

	\centering
	\begin{subfigure}{88mm}
    	\centering
        \input{classA}
        \caption{class A}
    \end{subfigure}%
	\begin{subfigure}{88mm}
    	\centering
        \input{classB}
        \caption{class B}
    \end{subfigure}
	\begin{subfigure}{88mm}
    	\centering
        \input{classC}
        \caption{class C}
    \end{subfigure}%
	\begin{subfigure}{88mm}
    	\centering
        \input{classD}
        \caption{class D}
    \end{subfigure}
    \caption{Example orbits from all four classes. Radial axis is the semi-major axis relative to the mean semi-major axis. Angular is the mean longitude in a frame rotating with the mean motion associated with the mean semi-major axis. To view the evolution of each of these cases, visit  \url{http://goo.gl/JgJ7wZ}}
    \label{fig:evolution-examples}
\end{figure*}

\onlfig{
\begin{figure*}
	\centering
	\input{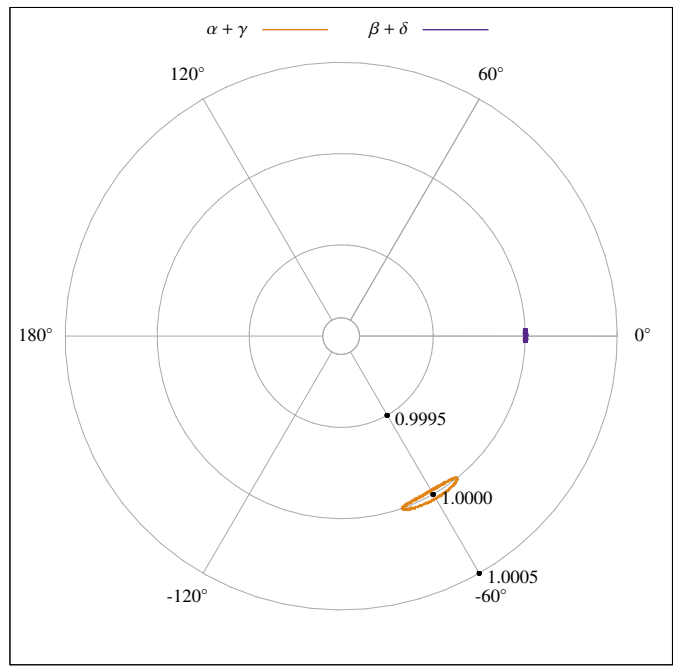}
	\caption{Single case of tadpole orbit. The fragments have the mass of Janus and Epimetheus. The plot shows the semi-major axis relative to the mean semi-major axis plotted against the mean longitude in a frame rotating with the mean motion of the mean semi-major axis.}
	\label{fig:tadpole}
\end{figure*}
}

\section{Conclusion and outlook}
We here investigated the possibility of a formation process of the Janus-Epimetheus system starting from a collision event that spawns four large fragments.
Additional debris was neglected.
We used numerical integration of the N-body problem to obtain the evolution of 8000 initial configurations.
We selected the configurations that led to the formation of a co-orbital system.
By analysing these, we established a common scenario that leads to the formation of co-orbital systems: After the fragments spread out on orbit, they re-collide and form a pair of co-orbiting bodies.

We investigated the circumstances of these secondary collisions and showed that the initial velocity has a high influence on the collision velocities.
At the same time, we found that lower initial velocities yield a higher probability of co-orbital formation.
This reinforces our assertion that for our assumed initial condition to be achieved, the prior collision event must have occurred at a low velocity.
It is unclear, however, whether the collision velocities observed in these simulations are low enough that a perfectly inelastic collision and joining of the bodies is a good approximation.
We assume that relative velocities below $58\mps$ allow the fragments to collide constructively.
Previous studies on the collisions of celestial bodies are not fully applicable to our system.
Most works dealt either with collisions of rocky bodies and/or of a small projectile hitting a large target.
Therefore it is necessary to study the collision behaviour of large low-density fluffy objects such as Janus and Epimetheus in more detail.
Studies in this area would help to ultimately decide the viability of our formation model.
We also found that there is a lower cut-off below which the formation of co-orbitals decreases significantly.
We attribute this to the inability of the fragments to escape their combined gravity when the kinetic energy is too low.
This imposes two constraints on the initial velocity:
It must be low enough to obtain reasonably low collision velocities.
At the same time, it must be high enough to allow the fragments to escape their mutual gravitational pull.
Our results indicate that this sweet spot is quite narrow, but indeed exists.

Another limiting feature of our physical model is the number of initial fragments adopted in the simulations. To strengthen the scenario proposed here, we performed two sets of simulations, one considering six, the other eight initial fragments. Considering initial velocities as those of simulation 3c, 100 simulations were performed for each case (six and eight fragments). The results showed an even higher rate of co-orbital generation (18-20\%).
Therefore, the scenario proposed here is somewhat robust.

Furthermore, we are aware that our initial conditions are not very natural.
Instead, we chose a simple configuration to constrain the otherwise quite large number of free parameters.
The main result of this study is therefore that the proposed formation model is possible.
We are confident to find co-orbital formation for a wider range of initial conditions building upon this result.
Ideally, a future work would include the initial collision as part of the simulation and use a better approximation for the secondary collision events. 
~\\
\begin{acknowledgements}
We thank Rafael Sfair for providing file storage for our results on short notice.
Othon C. Winter thanks FABESP (proc. 2011/08171-3) and CNPq.
The constructive comments and suggestions of the referee significantly improved this paper.
\end{acknowledgements}

\bibliographystyle{aa} \bibliography{paper}

\begin{thebibliography}{32}
\expandafter\ifx\csname natexlab\endcsname\relax\def\natexlab#1{#1}\fi

\bibitem[{Beaug{\'e} {et~al.}(2007)Beaug{\'e}, S{\'a}ndor, {\'E}rdi, \&
  Suli}]{beauge2007accretion}
Beaug{\'e}, C., S{\'a}ndor, Z., {\'E}rdi, B., \& Suli, A. 2007, Astronomy and
  Astrophysics, 463, 359

\bibitem[{Benz \& Asphaug(1999)}]{benz1999disruption}
Benz, W. \& Asphaug, E. 1999, Icarus, 142, 5

\bibitem[{Brown(1911)}]{brown1911}
Brown, E.~W. 1911, Monthly Notices of the Royal Astronomical Society, 71, 438

\bibitem[{Chambers(1999)}]{chambers1999mercury}
Chambers, J.~E. 1999, Monthly Notices of the Royal Astronomical Society, 304,
  793

\bibitem[{Chanut {et~al.}(2008)Chanut, Winter, \& Tsuchida}]{chanut2008nebular}
Chanut, T., Winter, O., \& Tsuchida, M. 2008, Astronomy and Astrophysics, 481,
  519

\bibitem[{Chanut {et~al.}(2013)Chanut, Winter, \& Tsuchida}]{chanut2013cavity}
Chanut, T., Winter, O., \& Tsuchida, M. 2013, Astronomy \& Astrophysics, 552,
  A66

\bibitem[{Charnoz {et~al.}(2010)Charnoz, Salmon, \&
  Crida}]{charnoz2010formation}
Charnoz, S., Salmon, J., \& Crida, A. 2010, Nature, 465, 752

\bibitem[{Chiang \& Lithwick(2005)}]{chiang2005neptune}
Chiang, E. \& Lithwick, Y. 2005, The Astrophysical Journal, 628, 520

\bibitem[{Colwell(1994)}]{colwell1994disruption}
Colwell, J. 1994, Planetary and Space Science, 42, 1139

\bibitem[{Cresswell \& Nelson(2006)}]{cresswell2006gasdrag}
Cresswell, P. \& Nelson, R. 2006, Astronomy and Astrophysics, 450, 833

\bibitem[{Crida \& Charnoz(2012)}]{crida2012formation}
Crida, A. \& Charnoz, S. 2012, Science, 338, 1196

\bibitem[{Dermott \& Murray(1981)}]{dermott1981dynamics}
Dermott, S.~F. \& Murray, C.~D. 1981, Icarus, 48, 12

\bibitem[{Fleming \& Hamilton(2000)}]{fleming2000}
Fleming, H.~J. \& Hamilton, D.~P. 2000, Icarus, 148, 479

\bibitem[{Giuppone {et~al.}(2012)Giuppone, Ben{\'i}tez-Llambay, \&
  Beaug{\'e}}]{giuppone2012cavity}
Giuppone, C., Ben{\'i}tez-Llambay, P., \& Beaug{\'e}, C. 2012, Monthly Notices
  of the Royal Astronomical Society, 421, 356

\bibitem[{Hadjidemetriou \& Voyatzis(2011)}]{hadjidemetriou2011gasdrag}
Hadjidemetriou, J.~D. \& Voyatzis, G. 2011, Celestial Mechanics and Dynamical
  Astronomy, 111, 179

\bibitem[{Harrington \& Seidelmann(1981)}]{harrington1981dynamics}
Harrington, R. \& Seidelmann, P. 1981, Icarus, 47, 97

\bibitem[{Izidoro {et~al.}(2010)Izidoro, Winter, \&
  Tsuchida}]{izidoro2010congential}
Izidoro, A., Winter, O., \& Tsuchida, M. 2010, Monthly Notices of the Royal
  Astronomical Society, 405, 2132

\bibitem[{Jacobson {et~al.}(2008)Jacobson, Spitale, Porco, Beurle, Cooper,
  Evans, \& Murray}]{jacobson2008orbits}
Jacobson, R., Spitale, J., Porco, C., {et~al.} 2008, The Astronomical Journal,
  135, 261

\bibitem[{Kary \& Lissauer(1995)}]{kary1995gasdrag}
Kary, D.~M. \& Lissauer, J.~J. 1995, Icarus, 117, 1

\bibitem[{Kortenkamp(2005)}]{kortenkamp2005capture}
Kortenkamp, S.~J. 2005, Icarus, 175, 409

\bibitem[{Kortenkamp \& Wetherill(2000)}]{kortenkamp2000collision}
Kortenkamp, S.~J. \& Wetherill, G.~W. 2000, Icarus, 143, 60

\bibitem[{Laughlin \& Chambers(2002)}]{laughlin2002collision}
Laughlin, G. \& Chambers, J.~E. 2002, The Astronomical Journal, 124, 592

\bibitem[{Leinhardt \& Stewart(2009)}]{leinhardt2009disruption}
Leinhardt, Z.~M. \& Stewart, S.~T. 2009, Icarus, 199, 542

\bibitem[{Murray(1994)}]{murray1994stability}
Murray, C.~D. 1994, Icarus, 112, 465

\bibitem[{Peale(1993)}]{peale1993trojans}
Peale, S. 1993, Icarus, 106, 308

\bibitem[{Renner \& Sicardy(2006)}]{renner2006elements}
Renner, S. \& Sicardy, B. 2006, Celestial Mechanics and Dynamical Astronomy,
  94, 237

\bibitem[{Salo \& Yoder(1988)}]{salo1988}
Salo, H. \& Yoder, C. 1988, Astronomy and Astrophysics, 205, 309

\bibitem[{Sicardy \& Dubois(2003)}]{sicardy2003}
Sicardy, B. \& Dubois, V. 2003, Celestial Mechanics and Dynamical Astronomy,
  86, 321

\bibitem[{Synnott {et~al.}(1981)Synnott, Peters, Smith, \&
  Morabito}]{synnott1981orbits}
Synnott, S.~P., Peters, C.~F., Smith, B.~A., \& Morabito, L.~A. 1981, Science,
  212, 191

\bibitem[{Thomas {et~al.}(2013)Thomas, Burns, Hedman, Helfenstein, Morrison,
  Tiscareno, \& Veverka}]{thomas2013details}
Thomas, P., Burns, J., Hedman, M., {et~al.} 2013, Icarus, 226, 999

\bibitem[{Yoder(1979)}]{yoder1979origins}
Yoder, C.~F. 1979, Icarus, 40, 341

\bibitem[{Yoder {et~al.}(1983)Yoder, Colombo, Synnott, \&
  Yoder}]{yoder1983theory}
Yoder, C.~F., Colombo, G., Synnott, S., \& Yoder, K. 1983, Icarus, 53, 431

\end{thebibliography}

\end{document}